# DS-GCNs: Connectome Classification Using Dynamic Spectral Graph Convolution Networks with Assistant Task Training

Xiaodan Xing [1, 2, *], Qingfeng Li [1, 4, **], Hao Wei [1, 5], Minqing Zhang [1, 6], Yiqiang Zhan [1], Xiang Sean Zhou [1], Zhong Xue [1], and Feng Shi [1]

[1] United Imaging Intelligence, Shanghai, China
[2] Shanghai Advanced Research Institute, Shanghai, China
[4] School of Biomedical Engineering, Southern Medical University, Guangdong, China
[5] School of Computer Science and Engineering, Central South University, Hunan, China
[6] College of Software Engineering, Southeast University, Jiangsu, China

* Corresponding author.

Xiaodan Xing

Shanghai Advanced Research Institute, Shanghai, China

xingxiaodan@sari.ac.cn

** Xiaodan Xing and Qingfeng Li contributed equally in this work.

# Abstract

Functional Connectivity (FC) matrices measure the regional interactions in the brain and have been widely used in neurological brain disease classification. However, a FC matrix is neither a natural image which contains shape and texture information, nor a vector of independent features, which renders the extracting of efficient features from matrices as a challenging problem. A brain network, also named as connectome, could form a graph structure naturally, the nodes of which are brain regions and the edges are interregional connectivity. Thus, in this study, we proposed novel graph convolutional networks (GCNs) to extract efficient disease-related features from FC matrices. Considering the time-dependent nature of brain activity, we computed dynamic FC matrices with sliding-windows and implemented a graph convolution based LSTM (long short time memory) layer to process dynamic graphs. Moreover, the demographics of patients were also used to guide the classification. However, unlike in conventional methods where personal information, *i.e.,* gender and age, were added as extra inputs, we argue that this kind of approach may not actually improve the classification performance, for such personal information given in dataset was usually balanced distributed. In this paper, we proposed to utilize the demographic information as extra outputs and to share parameters among three networks predicting subject status, gender and age, which serve as assistant tasks. We tested the performance of the proposed architecture in ADNI II dataset to classify Alzheimer's disease patients from normal controls. The classification accuracy, sensitivity and specificity reach 90.0%, 91.7% and 88.6% on ADNI II dataset.

# 1    Introduction

Neurological diseases usually lead to subtle structural brain abnormality. Functional MRI, which evaluates the brain activity by measuring the blood oxygen level over time, is thus a perfect tool to investigate possible brain functional changes in many neurological disorders. Considering the brain nature of functional integration and segregation, researchers assess the correlations among neuronal activities in order to analysis the brain function. A brain network forms a graph structure naturally. It contains a set of brain regions of interest (ROIs), known as nodes, and describes their connectivity, known as edges. In the context of fMRI, these edges are derived from the statistical dependencies between each pair of brain ROIs and are summarized in a symmetric matrix called functional connectivity (FC) matrix.

Note that unlike natural images that contains shape and texture information, the spatial locality of the entries of FC do not directly corresponds to the locality of brain networks. Thus, a difficult but important challenge in FC analysis is to extract efficient disease-related features. There are two major mainstreams of methods, one is to reshape FCs into vectors of features (Plis et al., 2014; Sen et al., 2016) and the other is to treat FCs as 2D images. However, both genres are problematic. A vectorized FC will loss the spatial information as discarding the topological structure of the matrix.

In brain network, graphical algorithms can be applied on FC data without loss of information. Graph kernel, which measures the inherent in the graph structure (Shervashidze, Schweitzer, van Leeuwen, Mehlhorn, & Borgwardt, 2011; Vishwanathan, Schraudolph, Kondor, & Borgwardt, 2010), is extensively used in many graph-based algorithms. Recently, researchers also applied graph kernel for neuro-imaging studies. For example, Jie et al. (Jie, Zhang, Wee, & Shen, 2014) used a graph-kernel-based approach to measure directly the topological similarity between connectivity networks. However, the computational complexity of graph kernel is intensive.

Graph convolution neural networks (GCN) allow an implementation of neural networks on graph structures. Neuroimaging pattern recognition applications include node classification and graph classification. Node classification assigns a pre-defined demographic graph for all subjects accompanied by a set of features. In the work of Parisot (Parisot et al., 2018), the feature of every individual, *i.e.*, every graph node, was a feature vector extracted from images, while the edges were calculated from the similarities between corresponding subjects. Graph classification treats each individual as an independent graph. For example, Ktena et al. (Ktena et al., 2018) proposed a Siamese graph convolution network to learn the similarity between brain networks. However, their works neglected the dynamic nature of brain activity, which can possibly improve the performance of classifiers.

Demographic information has been proved to be useful in many clinical studies. Researchers has reported a gender difference on cognitive functioning in brain (Halpern, 2012). Besides, the incidence of many neurological diseases correlates with the years of age (Braak & Braak, 1997; Butwicka & Gmitrowicz, 2010). Thus, gender and age information was used in many neurological disease classification studies. However, in small datasets, which is common in medical imaging scenario, we argue that the status of subjects and demographic information of subjects are not always strongly correlated because of sampling preference. For example, a major guideline of collecting data for diagnosis is to balance the demographic distribution in both patient group and healthy control group, making gender and age weakly correlates with diagnosis. In this case, involving gender and age directly in the prediction model may not help models to learn diagnosis better.

In our recent research presented at a conference (Xing et al., 2019), we proposed a novel graph convolution recurrent network for fMRI pattern recognition. Graphs are defined with time-varying edges, *i.e.*, the dynamic functional connections, and fixed nodes, which are characterized by the structural information retrieved from average fMRI. We demonstrate the potential of our method on connectome classification tasks. The proposed method can improve diagnosis accuracy on fMRI through three aspects: 1) A graph classification algorithm dedicated for functional connectivity data; 2) An LSTM architecture which could further extract temporal information relevant to diagnosis; 3) Multi-task settings which utilize demographic information as extra outputs for guiding the extraction of efficient features. In this work, we further extended the aforementioned study with more methodological details, validations and discussions.

## 2 Method

We introduce below the proposed dynamic spectral GCNs with assistant task training using dynamic functional connectivity matrices. **First**, the dynamic connectivity matrices based on correlations of BOLD signals from sliding-windows are defined as time-varying edges. Each node of the connectivity graph reflects an anatomical ROI. The feature on each node is defined by the volume of the corresponding ROI. **Second**, after defining the graph structure, a spectral graph convolution-based LSTM network is employed to extract information from the dynamic connectivity graphs. **Finally**, we make use of the demographic information as extra inputs, by adding two assistant networks with similar structure but different parameters, predicting gender and age. The feature maps from assistant networks were then weighted and combined with feature maps from main network, guiding the parameter training and optimizing the results.

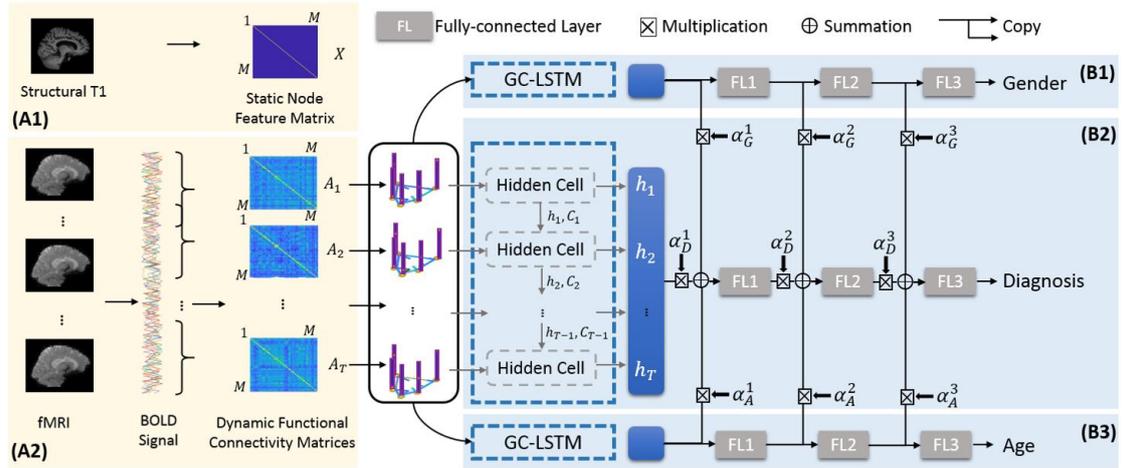

**Figure 1**. The schematic representation of the proposed DS-GCN with assistant task training for disease diagnosis using fMRI data. Time-varying edges are defined by dynamic functional connectivity (A1) and the static features of each node is defined by structural images (A2). Dynamic graphs are then processed by diagnosis network (B2) and two sub-networks (B1 & B3) for assistance.

### 2.1 Graph Construction

A graph $\mathcal{G}(\mathcal{V}, \mathcal{E})$ was defined by two matrices, *i.e.*, node feature matrix and adjacency matrix. Node feature matrix denoted as $X \in R^{M \times N}$, where $M$ is the number of nodes in the graph and $N$ denotes the number of features, describes the property of every node in the graph. And adjacency matrix $A \in R^{M \times M}$ represents the graph structure in a matrix form.

In our method, we obtain feature matrices from T1 images by assigning the volume of each ROI into corresponding diagonal entry. Thus, the feature matrix in our method is then denoted as $\{X|X \in R^{M \times M}, x_{ij} = \begin{cases} v_i, i = j \\ 0, i \neq j \end{cases}\}$, where $v_i$ is the brain volume of $i$-th ROI. Dynamic adjacency matrices are computed by a sliding window over the entire time series, and can be represented as $\{A_t|A_t \in R^{M \times M}, t = 1,2, \dots, T\}$. Here $T$ is the overall time points of dynamic adjacency matrices.

## 2.2 Graph Convolution LSTM

**Graph Convolution.** Conventional CNN performs spatial convolution on 2D or 3D images, which is reasonable because of the natural adjacent properties of neighboring pixels/voxels. However, when the input of neural network is a graph, such convolution operations should be designed specifically. In our study, we chose spectral graph convolution (Defferrard, Bresson, & Vandergheynst, 2016; Kipf & Welling, 2017), because of 1) the difficulty to match spatial local neighborhoods for a node and 2) the complete mathematical definition of spectral graph convolution. Consider a graph adjacency matrix $A$, the normalized graph Laplacian of $A$ is

$$L = I - D^{-\frac{1}{2}} A D^{-\frac{1}{2}}. \tag{1}$$

Here $D = diag(\sum_j a_{i,j})$ is the degree matrix of $A$. The eigenvectors $U$ of graph Laplacian form the Fourier bases of the graph. Thus, the Fourier transform of feature maps is $\hat{x} = U^T x$. Since the convolution operation on spatial domain is the multiplication on spectral domain, the graph convolution operation is defined as

$$g_\theta * x = U\big((U^T g_\theta) \odot (U^T x)\big), \tag{2}$$

where $g_\theta$ represents learnable parameters of the graph convolutional kernel. In our study, we used Chebyshev polynomial to approximate $U^T g_\theta$. This approximation enables spectral convolution to be spatial localized, and decreases the learning and computational complexity. Using Chebyshev polynomial, equation (2) can be approximated as

$$g_\theta * x = \sum_{j=0}^{K} \theta_j T_j(\tilde{L})x. \tag{3}$$

Here, Chebyshev polynomial is $T_k(x) = 2xT_{k-1}(x) - T_{k-2}(x)$ with $T_0(x) = 1$ and $T_1(x) = x$. $\tilde{L} = \frac{2}{\lambda_{max}}L - I$ is the scaled graph Laplacian, $\lambda_{max}$ denotes the largest eigenvalue of $L$ and $I$ is the identity matrix. $\theta_j$ is one of the learnable parameters. And this expression is $K$ localized because the $K$-th polynomial of the graph Laplacian only depends on maximum $K$-th nearest neighbor of the central node. For brain networks, every brain region densely connected with others, either weakly or strongly. Thus, we only need to consider the situation where $K = 1$. We further approximate $\lambda_{max} \approx 2$, and thus equation (3) is simplified as

$$g_\theta * x = \theta_0 x + \theta_1(L - I)x = \theta_0 x - \theta_1 D^{-\frac{1}{2}} A D^{-\frac{1}{2}} x \approx \theta'\left(\tilde{D}^{-\frac{1}{2}} \tilde{A} \tilde{D}^{-\frac{1}{2}}\right)x, \tag{4}$$

with $\tilde{A} = A + I$ and $\tilde{D} = diag(\sum_j \tilde{a}_{i,j})$. With equation (4), for input signal $X \in R^{M \times F_{in}}$, with $F_{in}$ input features for $N$ nodes and $F_{out}$ expected output features, we could define graph convolutional filter as follow:

$$Z = \tilde{D}^{-\frac{1}{2}} \tilde{A} \tilde{D}^{-\frac{1}{2}} X \Theta, \tag{5}$$

where $\Theta \in R^{F_{in} \times F_{out}}$ is a matrix of graph convolutional parameters and $Z \in R^{M \times F_{out}}$ is the output feature matrix.

**GC-LSTM.** Dynamic graphs require a recurrent structure to handle temporal information. In our paper, LSTM (Long Short-Tern memory) network is used, which could avoid the long-term dependency problem by recording the cell state in every time step. By replacing matrix multiplication in conventional LSTM with graph convolution, we could obtain below mathematical formulas which graph convolutional LSTM follows:

Forget Gate: $f_t = \sigma_f(\omega_{xf} * x_t + \omega_{hf} * h_{t-1} + \omega_{Cf} \odot C_{t-1} + b_f)$

Input Gate: $i_t = \sigma_i(\omega_{xi} * x_t + \omega_{hi} * h_{t-1} + \omega_{Ci} \odot C_{t-1} + b_i)$

Cell State: $C_t = f_t \odot C_{t-1} + i_t \odot tanh(\omega_{xc} * x_t + \omega_{hc} * h_{t-1} + b_c)$ $\qquad$ (5)

Output Gate: $o_t = \sigma_o(\omega_{xo} * x_t + \omega_{ho} * h_{t-1} + \omega_{Co} \odot C_t + b_o)$

Hidden State: $h_t = o \odot tanh(C_t)$.

Here, $*$ denotes the graph convolution operator. Forget gate decides what information are going to be discarded from the cell state, while input gate decides what new information to store. Cell state is updated after forget gate and input gate, and output gate calculates the output, which is then filtered by cell state and send to next time step. Details of the GC-LSTM cell is shown in Figure 2.

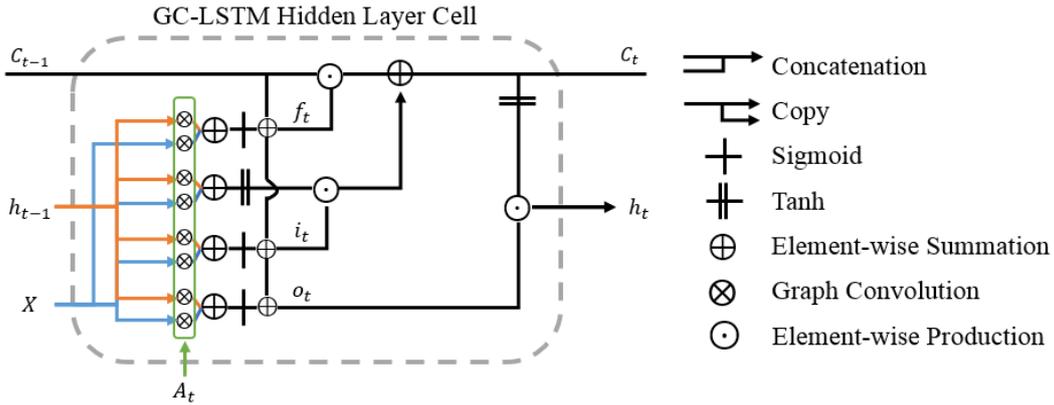

**Figure 2.** GC-LSTM hidden cell in detail. Each cell has two sources of input, the present ($A_t$ and $X$) and the recent past ($h_{t-1}$ and $C_{t-1}$). Output sequence $\{h_t | t = 1, 2, \dots T\}$ were then treated as feature maps for following layers.

## 2.3 Assistant Task Training

Gender and age provide important demographic information for disease prediction. Conventional deep learning classifiers added gender and age as additional features into the last fully connected layer. However, when adding demographic information as input features, the status and demographic information of subjects correlates in a statistical manner. For example, in ADNI datasets, gender and age are balanced among all groups, which makes these features correlates weakly with diagnosis labeling. In this situation, adding additional demographic features may not improve the classification performance. In addition, a balanced distribution in dataset does not imply a balanced distribution in real life: gender and age do affect the incidence of many neurological diseases (Hebert et al., 1995; Nebel et al., 2018). Thus, we propose to use the demographic information as extra outputs in our networks. This strategy not only could improve the classification performance of weakly correlated demographic features, but also could guide the parameter optimizing in diagnosis task.

However, conventional multitask settings adopt a hard representation-sharing manner, by which different tasks share same parameters in the beginning several convolutional layers of the network. In fact, it is difficult to decide which layers to be shared and which layers to be split among different tasks. Thus, our networks learn a linear combination of feature maps from different tasks to determine the shared representations by itself. We adopted this architecture from cross stitch networks (Misra, Shrivastava, Gupta, & Hebert, 2016), but our networks are task-centralized, *i.e.*, only the main network, diagnosis network, receives the weighted combination of feature maps from assistant networks.

Consider the feature maps $x_D^l$ from diagnosis network, $x_G^l$ from gender prediction network and $x_A^l$ from age prediction network in layer $l$, the linear combination of these feature maps are learnt and send into next layer in diagnosis network,

$$\tilde{x}_D = \alpha_D^l x_D^l + \alpha_G^l x_G^l + \alpha_A^l x_A^l \qquad (6)$$

Here, $\alpha^l$ is a learnable parameter in layer $l$, which determines the contribution of demographic assistance. The loss for these networks is the weighted combination of losses from three tasks. For diagnosis and gender prediction, cross entropy loss was used, while for age prediction, we used mean square error loss.

$$\mathcal{L} = \omega_D \mathcal{L}_D + \omega_G \mathcal{L}_G + \omega_A \mathcal{L}_A \qquad (7)$$

# 3 Experiments

We compared the proposed method with a number of state-of-the-art classification approaches to demonstrate the improvements brought by graph convolution LSTM and assistant task training.

**Data.** We evaluated the proposed method on public dataset ADNI[1] (Alzheimer's Disease Neuroimaging Initiative). Under a series of criteria[2], ADNI-2 dataset is split into 177 healthy controls and 115 patients with 1.5T T1-weighted structural images and functional images (subjects had their eyes open). From ADNI-2 dataset, we chose 212 for training and 80 for testing. Data distribution of both train and test dataset is shown in table 1.

|  | Training | | Testing | |
| --- | --- | --- | --- | --- |
|  | **AD** | **HC** | **AD** | **HC** |
| *N (Females)* | 82(42) | 130(69) | 36(15) | 44(24) |
| *Age (Mean ±SD)* | 76.2 ±6.0 | 74.6 ±7.1 | 74.9 ±8.3 | 74.9 ±6.2 |

**Table 1. Demographic information of ADNI II dataset.**

**Pre-Processing.** Pre-processions of both structural and functional images are under a standard pipeline. The pre-processing steps of data is shown in Figure 4. For structural images, we performed anterior commissure (AC) — posterior commissure (PC) correction and then resampled them into size $256 \times 256 \times 256$ with a resolution of $1 \times 1 \times 1 \text{ mm}^3$. Then, after intensity inhomogeneity correction by N3 algorithm, structural images were skull-stripped and registered into MNI space. Functional images were first slice time corrected by interpolation and motion corrected by a rigid body transformation on volumes, where mutual information was used as the cost function. Then, functional images were rigidly registrated to corresponding T1 images, and aligned to MNI space using the warping parameters of T1 to MNI space. Thus, functional images were demarcated into 116 ROIs by AAL (Automated Anatomical

---

[1] http://adni.loni.usc.edu/
[2] http://adni.loni.usc.edu/wp-content/themes/freshnews-dev-v2/documents/clinical/ADNI-2_Protocol.pdf

Labeling) template. Spatial filtering was applied with a Gaussian kernel with 4 mm FWHM (Full width at Half Maximum). BOLD signals were then temporally filtered using a band-pass filter between 0.01 and 0.1 Hz. Dynamic functional connectivity matrices were computed by a sliding window after four nuisance covariates were regressed out, including head motion parameters, global mean signal, white matter signal and cerebrospinal fluid (CSF) signal. Fisher-Z transformation was applied on all FC matrices.

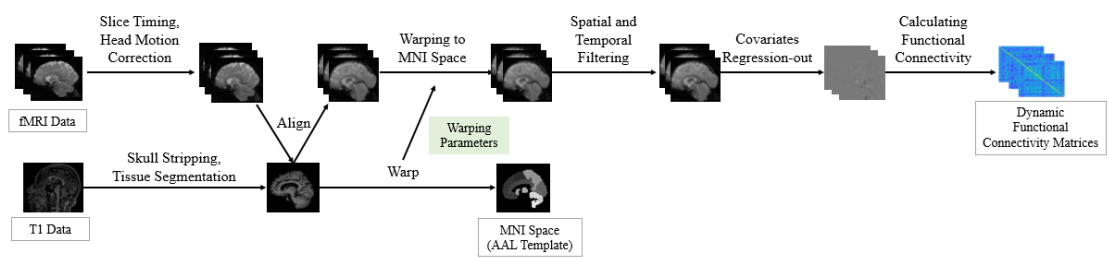

**Figure 4. Preprocessing steps of structural and functional data.** Pre-processions of T1 images include AC-PC correction, resampling, intensity correction, skull-stripping and registration. Pre-processions of fMRI include slice time correction, motion correction, registration, spatial and temporal filtering and nuisance covariates regression.

## 3.1 Baseline Models

We compared the proposed method with a number of state-of-the-art classification approaches on ADNI II dataset to demonstrate the improvements brought by graph convolution LSTM and assistant task training. We included 8 comparison methods in total, which compared two types of inputs, static FCs and dynamic FCs, and are shown in Figure 5. To assure a fair comparison, all methods were trained under super parameters (learning rate, batch size, etc.) which could optimize their performances on validation set.

**Static FCs.** 1) SVM: Static FCs and brain ROI volumes were reshaped as vectors of features and were then put into SVM with Gaussian kernel. 2) CNN: Static FCs and node feature matrices were treat as two channel images and put into VGG-16. 3) GCN: Static graphs were constructed by static FCs and node feature matrix, and then put into a graph convolution network with one graph convolution layer and three fully connected layers.

**Dynamic FCs.** 4) LSTM: Dynamic FCs were reshaped as a time-series of features, and then concatenated with static vectorized node feature matrices at every time point. 5) DS-GCN: No demographic information is used in this method. Dynamic graphs were constructed from functional connectivity matrices and structural information. The proposed network has one GC-LSTM layer followed by three fully connected layers. 6) DS-GCN with Demographics: In this method, demographic information, *i.e.,* gender and age, were used as extra inputs in the last fully connected layer. 7) DS-GCN with Hard Parameter Sharing: We compared hard parameter sharing, where different tasks share the same parameters in front layers of the network with the proposed algorithm. Networks were split for three tasks at the last fully-connected layer. 8) DS-GCN with Soft Parameter Sharing: Parameters in this method were initialized by the trained parameter in method (6).

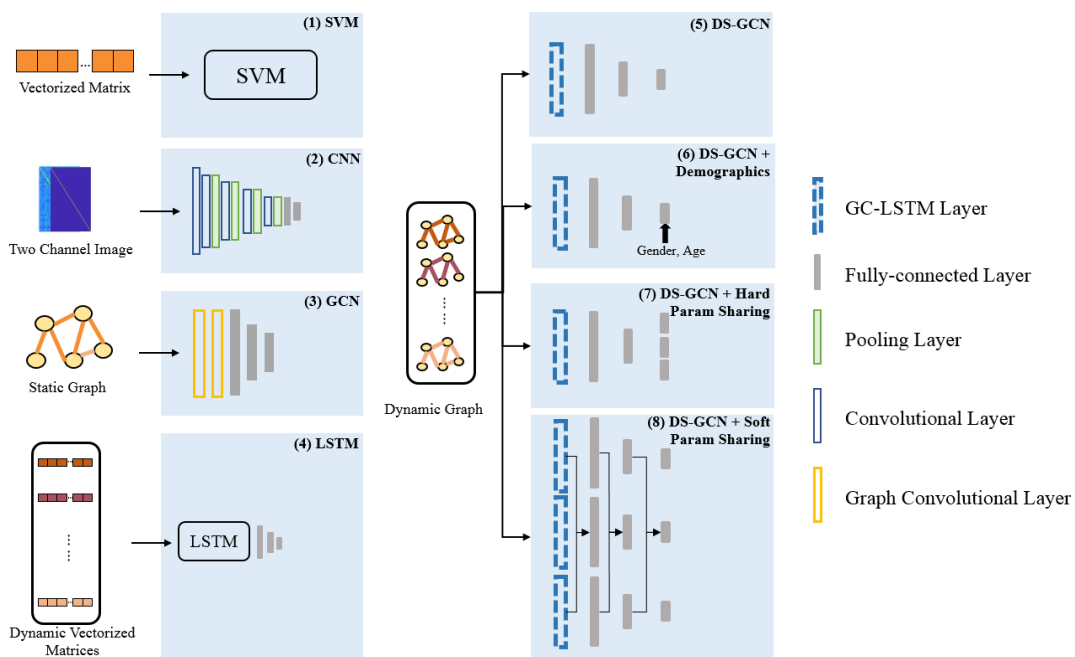

**Figure 5. Schematic representations of all methods for comparison.** It should be noted that for every layer in our comparison, batch normalizations and *ReLU* activations were contained.

## 3.2 Results

Classification results on ADNI II dataset is shown in TABLE 2. From the table one can observe that overall graph-based networks performed better than no-graph-based methods. In addition, dynamic connectivity outperformed static connectivity methods.

When we compared different settings of our algorithm, the performance by using assistant task training outperformed those without such a training strategy. The results indicates that dynamic connectivity with graph-based neural networks could fully exploit the information in fMRI connectivity analysis. Meanwhile, assistant task training may help improve the performance of classification.

| Inputs | Methods | Accuracy | Sensitivity | Specificity |
|---|---|---|---|---|
| **Static FCs** | SVM (Linear Kernel) | 68.8% | 72.2% | 65.9% |
| | GCN | 81.3% | 88.9% | 75.0% |
| | CNN | / | / | / |
| **Dynamic FCs** | RNN | 78.8% | 86.1% | 72.7% |
| | DS-GCN | 83.8% | 80.6% | 86.4% |
| | DS-GCN with Demographics | 86.3% | 88.9% | 84.1% |
| | DS-GCN with Hard Parameter Sharing | 87.5% | **91.7%** | 81.8% |
| | DS-GCN with Soft Parameter Sharing | **90.0%** | **91.7%** | **88.6%** |

**Table 2. Classification results of different algorithms on ADNI II dataset.**

# 4    Discussion

To further demonstrate the improvements in our model and to evaluate the assistant task training strategy, we compared the computational cost of all methods for comparison, computed the class activation map of our model and plotted the contribution from assistant tasks during training. The proposed model consume less storage and operations than other methods, and located bilateral hippocampus, right precuneus, right frontal middle cortex and left precentral cortex as class activated brain regions. Feature maps from gender and age prediction tasks have shown increasing contributions to diagnosis task through training.

## 4.1 Computational Cost

We compared the FLOPs (Floating Point Operations) (Hunger, 2005) and the physical size of aforementioned models. Comparing to baseline models without using demographic information, the learnable parameters are tripled in our methods with soft parameter sharing. Besides, to handling dynamic FC, we implemented LSTM structure into our network, also causing a considerable increase of parameter amount.

However, as is shown in Table 3, three times larger than baseline DS-GCN, the proposed method still consume less storage and operations than state-of-the-art methods, such as SVM, CNN and RNN. We attribute it to the computational simplicity of GCN. Consider an adjacency matrix $A^{116 \times 116}$ and a node feature matrix $X^{116 \times 116}$. Vectorizing both matrices as input, SVM requires at least 13456 parameters to assign a prediction. However, if we adopt graph convolution as described in equation (5), the dimension of parameters needed declines into $116 \times 1$. Graph convolution allows matrix calculation on functional connectivity matrices and thus considerably reduce the number of parameters. It is also worth noting that in GCN model, we implemented two graph convolution layers to assure a stable and acceptable performance, leading GCN model more complex than DS-GCN.

| Methods | Input Resolution | Size/kiB | FLOPs/kMAC |
|---|---|---|---|
| **SVM (Linear Kernel)** | $13572 \times 1 \times 1$ | $18.4 \times 10^2$ | 13.2 |
| **GCN** | $116 \times 116 \times 1$ | 64.4 | 22.2 |
| **CNN** | $116 \times 116 \times 2$ | $12.2 \times 10^3$ | $14.5 \times 10^6$ |
| **RNN** | $13572 \times 1 \times T$ | $52.5 \times 10^5$ | $9.3 \times 10^4$ |
| **DS-GCN** | $116 \times 116 \times T$ | 39.5 | 21.0 |
| **DS-GCN with Demographics** | $116 \times 116 \times T + 2$ | 39.5 | 21.0 |
| **DS-GCN with Multi-task** | $116 \times 116 \times T$ | 41.1 | 21.1 |
| **DS-GCN with Assistant task training** | $116 \times 116 \times T$ | 121.7 | 72.1 |

**Table 3. Parameter sizes of different models.** For ADNI-2 dataset, $T = 39$.

## 4.2 Network Explainability

Despite of the superior performance of graph convolutional models, the explainability of the model is also helpful, because graph structures cannot be classified easily by human intuition. Recently, an increasing number of researches has been proposed to study the inner working of graph convolutional networks(Baldassarre & Azizpour, 2019; Pope, Kolouri, Rostami, Martin, & Hoffmann, 2019). Gradient-guided Class Activation Mapping (Grad-CAM) is a widely used algorithm to interpret the decision making procedure of neural networks (Selvaraju et al., 2016). First, Grad-CAM calculates the gradient of $y_c$ (probability for class $c$, in this case, the probability for AD) with respect to feature maps $x_D^1$ output from GC-LSTM layer. Then, the importance weight $\omega_{c,k}$ of $k$-th node in this feature map is then

$$\omega_{c,k} = \frac{\partial y_c}{\partial x_{D,k}^1} \ .$$

The class activation map $M$ is feature map $x_D^1$ filtered by importance weight:

$$M_c = ReLU(\omega_c x_D^1).$$

Here, activation function ReLU makes heat map focusing on brain regions which only play positive parts in classification. The class activation value of all test subjects are averaged and shown in Figure 6 and 7.

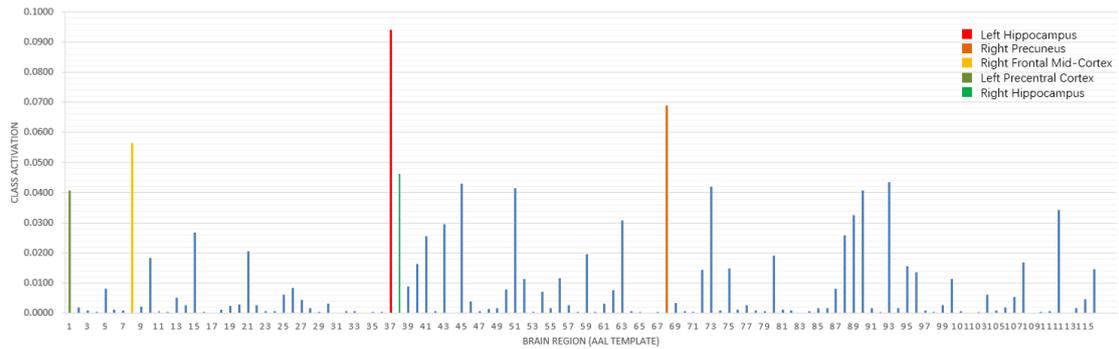

**Figure 6. A bar chart showing the average class activation value of all test subjects.** Top 5 activated regions are shown in different colors. Left hippocampus is reported as the most active region when our model diagnosis AD.

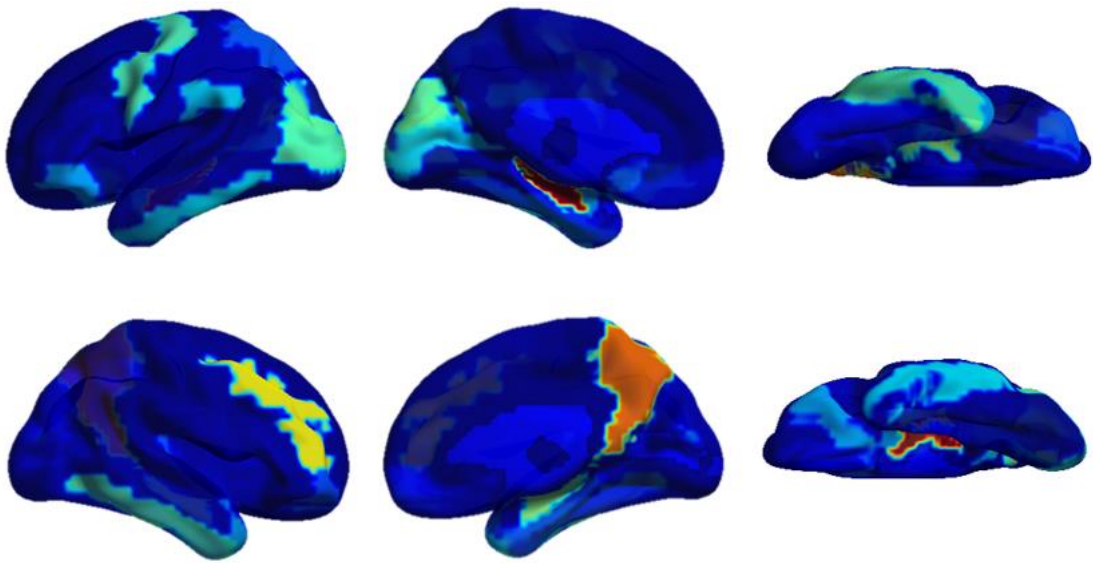

**Figure 7. The Alzheimer's disease activation map shown on a smoothed MNI 152 template.** This images is shown by BrainNetViewer (Xia, Wang, & He, 2013).

## 4.3 Assistant Task Training Analysis

**Synthetic Experiment.** To illustrate the effectiveness of assistant task learning, we reproduced the result of a synthetic experiment from (Caruana & De Sa, 1997). Here, we used two fully connected layer with activation function to approximate $(A + B)^2$. $A$ and $B$ are uniformly chosen from range $[-5,5]$ and are encoded into $2^{10}$ bins. Besides the binary codes of $A$ and $B$, another feature $(A - B)^2$ is also provided. However, $(A - B)^2$ weakly correlates with our target (the correlation reaches zero if $A$ and $B$ are "random" enough). This does not mean we should discard $(A - B)^2$.

By using $(A - B)^2$ as extra outputs, our simple network could easily generate the best prediction performance by learning to model sub-features $A$ and $B$. Figure 8 shows the network design and the result of this synthetic experiment.

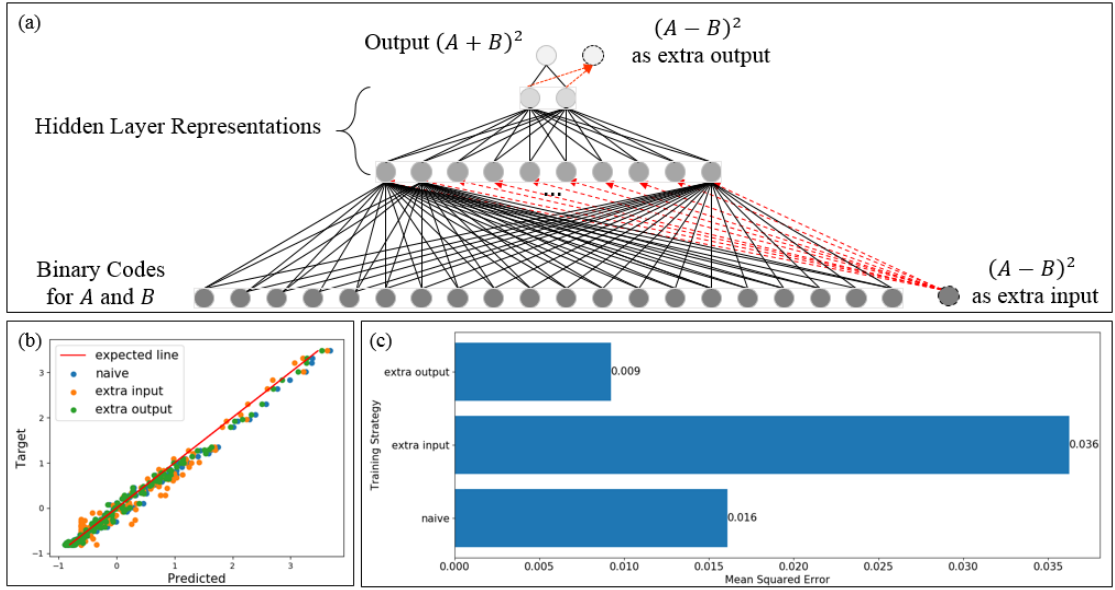

**Figure 8. Network design for synthetic experiment to illustrate the effectiveness of assistant task learning and its result** (both $(A - B)^2$ and $(A + B)^2$ are normalized). **(a)** The classification model is composed of three fully connected layers. Number of node here represents the dimension of features. **(b)** and **(c)** are the predicting results under three different learning strategy.

**Task Contribution.** The numeric values of α-s as well as the values of losses during training are plotted in Figure 9. To present the stable contribution from different tasks, we compared the updating values of alphas under different initializations. Experiment results show that using the training part of ADNI dataset, from which 43 subjects (20% of the training dataset) was selected for validation. Training procedure in all experiments used Adam (Kingma & Ba, 2015) with initial learning rate $10^{-3}$. For random initialization, the cross entropy loss of diagnosis and gender classification were between [0.6, 0.8], while the MSE loss of age prediction were between [0.07, 0.08]. Thus, to weight the losses from different tasks into the same scale, the weights of losses were then $\omega_D = \omega_G = 1, \omega_A = 10$. As seen in Eq. 6, $\alpha_D$ refers to the weight of the main tasks, $\alpha_G$ refers to the contribution from gender, and $\alpha_A$ refers to the contribution from age. These values reflect the influences of each factor to the final classification performance.

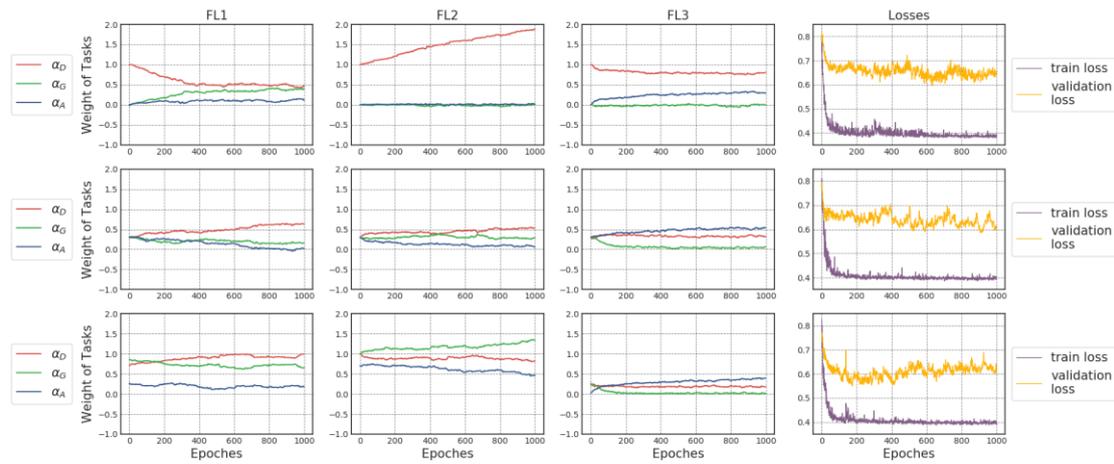

**Figure 9. Alpha values of different layers under different initializations.** Assistant task training shown in the first row was initialized with $\alpha_D = 1$ and $\alpha_G = \alpha_A = 0$, while in the second row was initialized with $\alpha_D = \alpha_G = \alpha_A = 0.33$. Figures of the third row show alpha values randomly initialized.

The results confirmed our hypothesis that the training of gender classification and age prediction may help generate stable and robust network feature maps for diagnosis tasks. Alphas in the last convolutional layers did not change as dramatically as the first two layers, indicating a weaker correlation among different tasks for deeper layers.

# 5    Conclusion

In this paper, we proposed a model specially to deal with dynamic functional connectivity. The novelty of this paper can be shown in three aspects. First, we proposed a specially designed graph construction algorithm, which could utilize both functional and structure images; Second, a spectral graph convolution based recurrent network is implemented to extract both functional and spatial information; Last, our model adopts a training strategy which utilizes demographic features as extra outputs, guiding the diagnosis network to train and focus.